\documentstyle[prl,aps,epsf]{revtex}
\begin{document}

\title{ Precise Asymptotics for a Random Walker's Maximum}
\author {Alain Comtet $^{1,2}$ and Satya N. Majumdar $^{1}$}
\address{
{\small $^1$Laboratoire de Physique Th\'eorique et Mod\`eles Statistiques,
        Universit\'e Paris-Sud. B\^at. 100. 91405 Orsay Cedex. France}\\
{\small $^2$Institut Henri Poincar\'e, 11 rue Pierre et Marie Curie, 75005 Paris, France}
}

\maketitle

\date{\today}

\begin{abstract}

We consider a discrete time random walk in one dimension. At each time step the walker
jumps by a random distance, independent from step to step, drawn from an arbitrary symmetric density 
function. We show that the expected positive maximum $E[M_n]$ of the walk up to $n$ steps
behaves asymptotically for large $n$ as, $\,$ $E[M_n]/\sigma=\sqrt{2n/\pi}+ \gamma +O(n^{-1/2})$, $\,$
where $\sigma^2$ is the variance of the step lengths. While the leading $\sqrt{n}$ behavior
is universal and easy to derive, the leading correction term turns out to be a nontrivial
constant $\gamma$. For the special case of uniform distribution over $[-1,1]$, Coffmann et. al.
recently computed $\gamma=-0.516068\dots$ by exactly enumerating a lengthy double series.
Here we present a closed exact formula for $\gamma$ valid for arbitrary symmetric distributions.
We also demonstrate how $\gamma$ appears in the thermodynamic limit as the leading behavior
of the difference variable $E[M_n]-E[|x_n|]$ where $x_n$ is the position of the walker
after $n$ steps. An application of these results to the equilibrium thermodynamics
of a Rouse polymer chain is pointed out. We also generalize our results to L\'evy 
walks.

\end{abstract}

\section{Introduction}

Brownian motion is perhaps one of the most widely studied subjects in classical physics.
A Brownian walker moves in continuous space and time which leads to a great simplification: one
can write down a simple differential equation, the famous diffusion equation, that governs
the time development of the probability density  of the position of the walker. Subsequently,
many other involved properties of the Brownian motion, such as its first-passage probability through
a given point, the distribution of the maximum displacement of the walker up to a given time
etc. can be calculated analytically relatively easily\cite{Redner}. In contrast, the related problem
of a random walker that hops only at discrete time steps in a continuous space is not
so straightforward, even in one dimension\cite{Feller,Hughes}. A classic example of such a walk can be found in 
bacterial
chemotaxis, where a bacteria, in search of food, jumps from one position to another at discrete time
steps\cite{Koshland}. Another famous example of such a walk occurs in the Rouse model
of a polymer chain that consists of monomers or beads connected by 
harmonic springs\cite{Rouse}. Many other examples can be found in Refs. \cite{MS,BKZ,BG}. While the 
asymptotic properties of this discrete hopper, after a sufficiently 
large number of steps and given that the variance in step sizes is finite, are correctly 
described
by the continuous time diffusion equation\cite{BG}, there are many interesting finite size effects
that can not be captured by the diffusion equation. The difficulty that arises
in dealing with a finite number of steps is due to the fact that 
the probability density of the random hopper usually satisfies an integral equation
which is technically much harder to solve than a differential equation. In this paper,
we study analytically one such finite size effect, namely the behavior of the expected
maximum position of a discrete time hopper in one dimension. We will show that even
this relatively simple problem has rather interesting finite size behavior.   

We consider a discrete time random walker moving on a continuous line. The
position $x_n$ of the walker after $n$ steps evolves for $n\ge 1$ via, 
\begin{equation}
x_n = x_{n-1}+ \xi_n
\label{evol1}
\end{equation}
starting at $x_0=0$, where the step lengths $\xi_n$'s are independent and identically distributed
(i.i.d) random variables with zero mean and each drawn from the same probability distribution, 
${\rm Prob}(\xi_n\le x)=\int_{-\infty}^x f(y)\,dy$,  $f(x)$ being the normalized symmetric probability 
density. 
Let $M_n$ denote the positive maximum of the random walk up to $n$ steps,
\begin{equation}
M_n = {\rm max} (0, x_1, x_2, \dots, x_n).
\label{max1}
\end{equation}
We are interested in the asymptotic large $n$ behavior of the expected maximum $E(M_n)$. 

This question recently arose in the context of a packing problem in two dimensions 
where $n$ rectangles of variable sizes
are packed in a semi-infinite strip of width one\cite{CS,CFFH}.
It was shown in Ref. \cite{CFFH} that for the special case of the uniform jump distribution,
$f(x)=1/2$ for $-1\le x \le 1$ and $f(x)=0$ outside, for large $n$, 
\begin{equation}
E[M_n] = \sqrt{\frac{2n}{3\pi}} -0.297952\dots + O(n^{-1/2}).
\label{flaj1}
\end{equation}
The leading $\sqrt{n}$ behavior is easy to understand and can be derived from the corresponding
behavior of a continuous time Brownian motion after a suitable rescaling\cite{CFFH}. However,
the leading finite size correction term turns out to be a nontrivial constant $c=-0.297952\dots$ that 
was computed in Ref. \cite{CFFH} by enumerating a somewhat awkward double series obtained after
a lengthy calculation. This constant $c$ characterizing the leading finite size behavior is
nonuniversal and is expected to depend on the details of the probability density $f(x)$ of the
noise. A natural question is: can one calculate this constant for arbitrary density function
$f(x)$? In this paper we provide an exact formula for this constant $c$ valid for arbitrary
symmetric $f(x)$. 

Our results are twofold. First we consider the class of density function $f(x)$ that has a
finite second moment, $\sigma^2= \int_{\infty}^{\infty} x^2 f(x) dx$. Then $\sigma$ denotes
the characteristic length of a single jump. Since, $E[M_n]$ has the dimension of length, it is
preferable to consider the dimensionless variable $E[M_n]/\sigma$. We show that for large $n$
\begin{equation}
\frac{E(M_n)}{\sigma}= \sqrt{\frac{2 n}{\pi}} + \gamma + O\left(\frac{1}{\sqrt{n}}\right).
\label{asymp1}
\end{equation}
The leading $\sqrt{n}$ behavior is universal (does not depend on the details of
the density function $f(x)$) and easy to compute by appropriately rescaling 
the continuous time Brownian result. Our main new result is to obtain an
exact expression for the nonuniversal constant $\gamma$. Our result is 
best expressed in terms of the characteristic function, 
\begin{equation}
{\hat f}(k)= \int_{-\infty}^{\infty} f(x)\, e^{ikx}\,dx.   
\label{chf1}
\end{equation}
For density functions with a finite second moment, i.e., when ${\hat f}(k)=1-\sigma^2 k^2/2+ O(k^4)$
as $k\to 0$, we show that
\begin{equation}
\gamma= \frac{1}{\pi \sqrt{2}} \int_0^{\infty} \frac{dk}{k^2} 
\ln {\left[ \frac{1-{\hat f}\left(\frac{\sqrt{2}}{\sigma}k\right)}{k^2} \right]}.
\label{c1}
\end{equation}
We prove in appendix-A that $\gamma<0$ for arbitrary $f(x)$, a fact not apriori obvious. Let us quote
a few examples where the integral in Eq. (\ref{c1}) can be performed explicitly,
\begin{eqnarray}
f(x)&=& \frac{1}{2}[\delta(x+1)+\delta(x-1)]\,\, \Rightarrow  \,\, \gamma=-1/2=-0.5 \\
f(x)&=& \frac{1}{2}e^{-|x|} \,\, \Rightarrow  \,\, \gamma=-1/\sqrt{2}=-0.70710\dots \\
f(x)&=& \frac{a^2}{2} |x|\,e^{-a|x|}\,\, \Rightarrow  \,\, \gamma=-(2\sqrt{3}-1)/\sqrt{6}=-1.00597\dots \\
f(x)&=& \frac{1}{\sigma \sqrt{2\pi}} e^{-x^2/2\sigma^2} \,\, \Rightarrow  \,\, \gamma=\zeta(1/2)/\sqrt{2\pi}=-0.58259\dots.
\label{special1}
\end{eqnarray}
where $\zeta(z)$ is the Riemann zeta function (analytically continued for $z<1$). For the uniform distribution over
$[-1,1]$, our exact formula in Eq. (\ref{c1}) reproduces very simply the result obtained in Ref. \cite{CFFH}. In
this case, using ${\hat 
f}(k)= \frac{\sin k}{k}$ and $\sigma=1/\sqrt{3}$ in Eq. (\ref{c1}) one gets
\begin{equation}
\gamma= \frac{1}{\pi \sqrt{2}}\int_0^{\infty} \frac{dk}{k^2} \ln {\left[ 
\frac{1-\frac{\sin(\sqrt{6}k)}{\sqrt{6}k}}{k^2}\right]}
=-0.516068\dots 
\label{cuni}
\end{equation}
where the integral was performed using Mathematica. Note that the constant $c=-0.297952\dots$ in Eq. (\ref{flaj1})
is simply $c=\gamma \sigma$ with $\sigma=1/\sqrt{3}$ for the uniform distribution and $\gamma=-0.516068\dots$
given in Eq. (\ref{cuni}).

We have also generalized our results to the case of L\'evy flights where the second moment diverges and
one has ${\hat f}(k) = 1- |ak|^{\mu}+O(k^2)$ as $k\to 0$ with $1< \mu \le 2$\cite{Bertoin}. Note that $a$ has the
dimension of length. The probability
density $f(x)$ of the step lengths has an algebraic tail for large $|x|$, $f(x)\sim |x|^{-1-\mu}$.
In this case we show that for large $n$, the dimensionless expected maximum behaves as 
\begin{equation}
\frac{E(M_n)}{a}= \frac{\mu}{\pi} \,\Gamma\left(1-\frac{1}{\mu}\right) n^{1/\mu} + \gamma 
+O(n^{1/\mu-1}). 
\label{asymp2}
\end{equation}
The leading term is again shown to be universal. We show that the leading finite size correction term 
is again a constant
given by
\begin{equation}
\gamma= \frac{1}{\pi}\int_0^{\infty} \frac{dk}{k^2} \ln {\left[\frac{1-{\hat f}\left(\frac{k}{a}\right)}
{k^\mu}\right]}.
\label{c2}
\end{equation}
For example, for the case ${\hat f}(k)= \exp[-|ak|^\mu]$ with $1< \mu\le 2$, we obtain
\begin{equation}
\gamma= \frac{1}{\pi}\int_0^{\infty} \frac{dk}{k^2} \ln 
{\left[\frac{1-e^{-k^{\mu}}}{k^{\mu}}\right]}=\frac{\zeta(1/\mu)}{(2\pi)^{1/\mu} \sin (\pi/{2\mu})}.
\label{special2}
\end{equation}
The evaluation of the integral is presented in appendix-B.

An interesting fact about the constant $\gamma$ is that even though it characterises
the leading finite size correction to the expected maximum, it actually shows up even
in the thermodynamic limit $n\to \infty$ provided one looks at a behavior of a
suitably defined quantity as follows. Let $|x_n|$ denote
the absolute value of the position of the walker after $n$ steps. The distribution of $x_n$, for
arbitrary density $f(x)$, can be computed relatively easily (see section-IV) and 
hence one can calculate $E[|x_n|]$.
We focus here on the case when the variance $\sigma^2=\int_{-\infty}^{\infty} x^2 f(x)dx$
as well as the fourth moment $\mu_4=\int_{-\infty}^{\infty} x^4 f(x) dx$ of the jump distribution
are finite. In that
case, one can show that for large $n$,
\begin{equation}
\frac{E[|x_n|]}{\sigma} = \sqrt{\frac{2 n}{\pi}}- 
\frac{1}{12\sqrt{2\pi}}\left(\frac{\mu_4}{\sigma^4}-3\right)\frac{1}{\sqrt{n}} + O(n^{-3/2}).
\label{xn1}
\end{equation}
Thus the leading term of $E[|x_n|]/\sigma$ for large $n$ is exactly the same as that of 
the expected maximum $E[M_n]/\sigma$ in
Eq. (\ref{asymp1}). However, unlike in the case of the maximum in Eq. (\ref{asymp1}), the leading finite size 
correction term in Eq. (\ref{xn1}) is of $O(n^{-1/2})$ and not a constant. 
Using Eqs. (\ref{asymp1}) and (\ref{xn1}) one then gets
\begin{equation}
\frac{E[M_n]-E[|x_n|]}{\sigma} = \gamma + O(n^{-1/2}).
\label{xn2}
\end{equation}
Thus the difference between the expected positive maximum up to $n$ steps and the absolute
value of the expected final position of the walker after $n$ steps, in units of $\sigma$, tends to
a negative constant $\gamma$ in the thermodynamic limit $n\to \infty$, a fact that is not  
apriori obvious.     

We end this section by mentioning a simple physical application of the results above. Let us consider
the simplest model of a polymer chain namely the Rouse model\cite{Rouse} where the monomers
are connected by harmonic springs. A configuration of the chain consisting of $n$ monomers 
is specified by the position vectors $\{\vec r_i\}$ of the monomers with $i=0,1,2 \dots, n$.
We assume that one end of the chain is grafted at the origin, ${\vec r_0}=0$ while the 
other end is free. We assume that the chain is at thermal equilibrium so that the probability
of any given chain configuration is given by its Boltzmann weight,
\begin{equation} 
P\left[\{ {\vec r_i}\}\right] = \frac{1}{Z_n}\exp\left[-\frac{\beta \kappa}{2}\sum_{i=1}^{n}\left(\vec r_i 
-\vec r_{i-1}\right)^2\right],
\label{eqcon1}
\end{equation}
where $Z_n$ is the partition function, $\beta=1/{k_B T}$ is the inverse temperature and the spring 
constant $\kappa$ characterises the harmonic coupling between neighbouring monomers. Let us now
look at the components of the position vectors along any particular direction, say $\{ x_i\}$, which
can also be thought of as a one dimensional Rouse chain. The equilibrium weight in Eq. (\ref{eqcon1})
indicates that the difference in position between the $i$-th and $(i-1)$-th monomer can be represented
by a noise
\begin{equation}
x_i = x_{i-1} + \xi_i,
\label{eqcon2}
\end{equation}
where $\xi_i$'s are independent and Gaussian distributed , 
$f(\xi)= e^{-\xi^2/{2\sigma^2}}/{\sigma \sqrt{2\pi}}$ where $\sigma^2=1/{\beta \kappa}$. Thus $M_n$ in Eq. 
(\ref{max1}) refers to the
maximum displacement of the polymer chain along $x$ direction and $x_n$ denotes the $x$ coordinate 
of the end point of the chain (see Fig. 1). 
\begin{figure}[htbp]
\epsfxsize=8cm
\centerline{\epsfbox{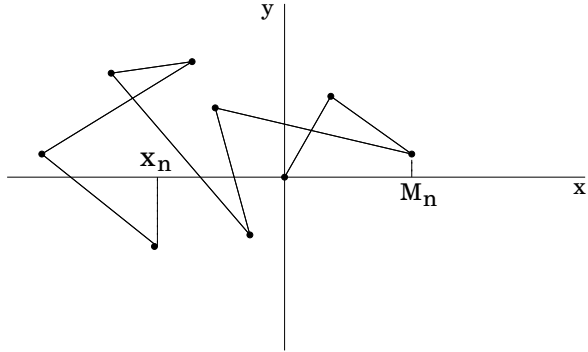}}
\caption{A typical configuration of the Rouse chain in $2$-dimensions. $M_n$ denotes the
positive maximum along the $x$ direction and $x_n$ denotes the $x$ co-ordinate of the position of the 
end point.}
\label{fig:polymer1}
\end{figure}

Thus, using the result in Eq. (\ref{special1}) in Eq. 
(\ref{xn2})
we find that in the limit of a very long chain ($n\to \infty$) at thermal equilibrium, the difference 
between the expected maximum displacement of the chain along a particular direction (say the $x$ direction)   
and the absolute end to end displacement along the same direction tends to a nontrivial constant
\begin{equation}
\frac{E[M_n]-E[|x_n|]}{\sigma} \to \frac{\zeta(1/2)}{\sqrt{2\pi}}= -0.58259\dots.
\label{xng}
\end{equation}
Note that in the context of the Rouse chain, the expectation $E$ means a thermal
equilibrium average over the Bolzmann weight in Eq. (\ref{eqcon1}).
The facts that (a) the difference approaches a constant and (b) that too, a negative constant,
are not apriori obvious for the Rouse chain.

The rest of the paper is organized as follows. In Section II, we set up the basic integral equation
for the distribution of the maximum and provide an exact solution in Section II-A for the
special case when the jump density is exponential. 
Section III deals
with the general jump distribution where we present the Pollaczek-Spitzer 
formula and extract the
finite size correction term exactly from an asymptotic expansion of this
formula.    
These results are generalized to L\'evy processses in Section III-B. 
In Section-IV we calculate the expected value of the absolute position
of the end point and demonstrate how the constant $\gamma$ shows up
in the thermodynamic limit. 
We conclude with a summary
and open problems. Explicit derivations of some of the formulae and integrations are relegated
to the two appendices.

\section{An Integral Equation for the distribution of the maximum}

In this section we set up an integral equation satisfied by the distribution
of the maximum of a random walk for arbitrary symmetric jump distribution. 
We consider a random walk starting at $x_0=0$ at step $n=0$ and evolving via Eq. (\ref{evol1}) 
where the noise $\xi_n$'s
are i.i.d. variables drawn from the common symmetric distribution
${\rm Prob}(\xi_n\le x)=\int_{-\infty}^x f(y)\,dy$. We would like to
compute the distribution of the maximum $M_n={\rm max}(0,x_1, x_2, \ldots, x_n)$ 
up to $n$ steps, i.e., ${\rm Prob}[M_n\le y]$. To derive this, we first
define $Q_n(x,y)$ as the probability that, starting at $x_0=x$, the maximum of the walk up to $n$ 
steps is less than or equal to $y$. Evidently, ${\rm Prob}[M_n\le y]=Q_n(0,y)$. 
Consider the first step where the particle jumps by an amount $x_1-x$ which occurs with
a probability density $f(x_1-x)$ (see Fig. 2). 
\begin{figure}[htbp]
\epsfxsize=8cm
\centerline{\epsfbox{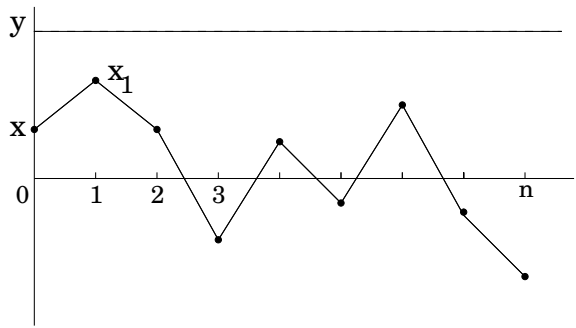}}
\caption{A random walker, starting at $x$ at $n=0$, makes a flight to $x_1\le y$ at step
$n=1$.}
\label{fig:rwfig1}
\end{figure}

It follows, using the Markov property
of the walk, that $Q_n(x,y)$ satisfies the following recursion relation,
\begin{equation}
Q_n(x,y)= \int_{-\infty}^{y} Q_{n-1}(x_1,y)\, f(x_1-x)\,dx_1
\label{recur1}
\end{equation}
with the initial condition $Q_0(x,y)=\theta(y-x)$ where $\theta(z)$ is the Heaviside
step function. Due to the translational invariance, it is also clear that $Q_n(x,y)$
depends only on the difference $z=y-x$, i.e., $Q_n(x,y)=q_n(z=y-x)$ where $z\ge 0$. 
Making the change of variable, $z'=y-x_1$ and using the translation invariance, the recursion
in Eq. (\ref{recur1}) becomes simpler,
\begin{equation}
q_n(z)= \int_0^{\infty} q_{n-1}(z') f(z-z')\, dz',
\label{recur2}
\end{equation}
valid for all $z\ge 0$ and starting with $q_0(z)=\theta(z)$. Thus, if one finds 
the solution $q_n(z)$ of Eq. (\ref{recur2}), then the distribution
of the maximum is just ${\rm Prob}[M_n\le y]=Q_n(0,y)=q_n(y)$. The
density of the maximum is $q_n'(y)=dq_n/dy$. Hence, the expected
maximum is $E[M_n]=\int_0^{\infty} q_n'(y) y\, dy$, the quantity
we are after.

The generating function, ${\tilde q}(z,s)=\sum_{n=1}^{\infty} q_n(z) s^n$ then satisfies
an integral equation     
\begin{equation}
{\tilde q}(z,s) = s\, \int_0^{\infty} {\tilde q} (z',s) f(z-z') dz' + s\, \int_0^{\infty} f(z-z')dz',
\label{int1}
\end{equation}
valid for all $z\ge 0$.
This integral equation is an inhomogeneous Wiener-Hopf equation\cite{MF} and in general, for arbitrary
kernel $f(z-z')$, it is very hard to solve this integral equation. The main source
of difficulty is the fact that the limits of the integral on the right hand side of
Eq. (\ref{int1}) are $0$ and $\infty$, as opposed to 
say $-\infty$ and $\infty$\cite{MF,Feller}. 
However, when
the kernel $f(z-z')$ is a normalized probability density function we may use
the Pollaczek-Spitzer
formula \cite{Pollaczek,Spitzer2}, to which we will come back
to in section III. But, before that, it is instructive to solve Eq. (\ref{int1})
explicitly for special cases, whenever possible. In Section II-A, we solve
Eq. (\ref{int1}) explicitly for the exponential density function.

\subsection{ Exponential density function: An exactly solvable case} 

For the exponential density functon, $f(z)= \frac{1}{2} e^{-|z|}$, one can 
obtain an exact solution of Eq. (\ref{int1}). We first assume that the
integral Eq. (\ref{int1}) is valid for all $-\infty \le z\le \infty$, though
we are interested only in the solution for $z\ge 0$. Next, we
note the identity, $f''(z)= f(z)-\delta(z)$, where $f''(z)=d^2f/dz^2$.
Differenting twice Eq. (\ref{int1}) and using the above identity one 
readily converts the integral equation into the following differential
equation,
\begin{equation}
\frac{d^2 {\tilde q}}{dz^2}= [1-s\theta(z)]{\tilde q} - s \theta(z).
\label{diff1}
\end{equation}
For $z\ge 0$, the general solution is readily obtained,
\begin{equation}
{\tilde q}(z,s) = \frac{s}{1-s} + A(s) e^{-\sqrt{1-s}\, z} + B(s) e^{\sqrt{1-s}\, z},
\label{diff2}
\end{equation}
where $A(s)$ and $B(s)$ are two arbitrary constants (independent of $z$). Now, in
the limit $z\to \infty$, $q_n(z)\to 1$ for all $n$, since the probability that
the particle, starting at $z\to \infty$, will not cross $0$ up to any finite step $n$
is $1$. Thus, one expects that as $z\to \infty$, ${\tilde q}(z,s)\to s/(1-s)$. Using
this boundary condition in Eq. (\ref{diff2}), one gets $B(s)=0$. The other
constant $A(s)$ will be fixed by the matching conditions at $z=0$. 

Now, for $z\le 0$, the solution of Eq. (\ref{diff1}) is given by
\begin{equation}
 {\tilde q}(z,s) = C(s)\, e^{z},
\label{diff3}
\end{equation}
where we have used the boundary condition, ${\tilde q}(z\to -\infty, s)\to 0$. Now,
we are ready to match the solution in Eq. (\ref{diff2}) for $z\ge 0$ with that
in Eq. (\ref{diff3}) for $z\le 0$. The continuity of the solution at $z=0$ and
also the continuity of the first derivative at $z=0$ fixes the two constants 
$A(s)$ and $C(s)$ uniquely. We get,
\begin{equation}
A(s)= -\frac{1-\sqrt{1-s}}{1-s}; \quad\quad\quad\, C(s)= \frac{1-\sqrt{1-s}}{\sqrt{1-s}}.
\label{const}
\end{equation}
Thus, for $z>0$, the exact solution of Eq. (\ref{int1}) is given by
\begin{equation}
{\tilde q}(z,s) = \frac{s}{1-s} -\frac{1-\sqrt{1-s}}{1-s}\, e^{-\sqrt{1-s}\,z}.
\label{sol1}
\end{equation}
The probability density for the maximum then has the generating function,
\begin{equation}
\sum_{n=1}^{\infty} q_n'(z) s^n = \frac{1-\sqrt{1-s}}{\sqrt{1-s}} e^{-\sqrt{1-s}\,z}.
\label{maxden1}
\end{equation} 
Subsequently, the generating function for the expected maximum is given by
\begin{equation}
\sum_{n=1}^{\infty} E[M_n] s^n = \sum_{n=1}^{\infty} s^n \int_0^{\infty} q_n'(z) z dz = 
\frac{1}{(1-s)^{3/2}}- \frac{1}{1-s}.
\label{maxgen1}
\end{equation}
Expanding the right hand side of Eq. (\ref{maxgen1}) in powers of $n$, we get 
\begin{equation}
E[M_n]= -1 + \frac{2}{\sqrt{\pi}} \frac{\Gamma(n+3/2)}{\Gamma(n+1)},
\label{maxav1}
\end{equation}
where $\Gamma(x)$ is the Gamma function. Note that the result in Eq. (\ref{maxav1}) is
valid for all $n\ge 0$. The variance of the step lengths for the exponential density
is given by, $\sigma^2 =\int_{-\infty}^{\infty} z^2 e^{-|z|}dz/2= 2$. 
Hence, the
dimensionless number $E[M_n]/\sigma$ is given by
\begin{eqnarray}
\frac{E[M_n]}{\sigma}&=& -\frac{1}{\sqrt{2}}+ \sqrt{\frac{2}{\pi}}\, 
\frac{\Gamma(n+3/2)}{\Gamma(n+1)}\nonumber \\
&= & \sqrt{\frac{2n}{\pi}} -\frac{1}{\sqrt{2}} + O(n^{-1/2}),
\label{sol2}
\end{eqnarray}    
where we have made the asymptotic expansion for large $n$ in the second line. Thus, the
result in Eq. (\ref{sol2}) is of the general asymptotic form as in Eq. (\ref{asymp1})
with the nontrivial constant 
\begin{equation}
\gamma= - \frac{1}{\sqrt{2}}.
\label{sol3}
\end{equation} 

\section{ The General Case: asymptotic expansion of Pollaczek-Spitzer formula}

In this section, we derive the exact asymptotic behavior in Eq. (\ref{asymp1}) for
the expected maximum for an arbitrary, symmetric jump density function $f(z)$. As
mentioned in the previous section, the solution of the integral equation Eq. (\ref{int1})
is hard to obtain analytically for an arbitrary kernel $f(z-z')$. However, 
when $f(z)$  is a  probability density function, Pollaczek derived
a general formula giving the Laplace transform of the probability density of 
ordered partial sums of random independent variables\cite{Pollaczek}.  
In the special case of 
the distribution of the maximum, this formula was
rederived by Spitzer \cite{Spitzer2} by a combinatorial approach.
In principle, this solves the problem. However, extracting the precise asymptotic behavior
of the first moment of the maximum is still nontrivial and this is what is precisely
achieved in this section. Consider the following Laplace transform,
\begin{equation}
E\left[e^{-\rho M_n}\right] = \int_0^{\infty} e^{-\rho z} q_n'(z) dz,
\label{spit1}
\end{equation}
where $q_n'(z)=dq_n/dz$ is the probability density of the maximum
and $q_n(z)$ satisfies the recursion relation in Eq. (\ref{recur2}). The
Pollaczek-Spitzer formula for the generating function of the above Laplace 
transform\cite{Spitzer1} reads
\begin{equation}
\sum_{n=0}^{\infty} s^n E\left[e^{-\rho M_n}\right] = \frac{1}{\sqrt{1-s}}\,\phi(s,\rho);\quad\quad {\rm 
where}\quad\, \phi(s,\rho)= \exp\left[-\frac{\rho}{\pi}
\int_0^{\infty} \frac{\ln \left(1-s {\hat f}(k)\right)}{\rho^2+k^2}\, dk \right],  
\label{spit2}
\end{equation}
where $0\le s\le 1$ and $f(z)$ is an arbitrary symmetric normalized density 
function. 
In Eq. (\ref{spit2}), ${\hat f}(k)= \int_{-\infty}^{\infty} f(z) e^{ikz}\,dz$ is
the Fourier transform of $f(z)$. 

The generating function for the expected maximum
can then be obtained by differentiation,
\begin{equation}
h(s) = \sum_{n=0}^{\infty} s^n E[M_n] = -\frac{1}{\sqrt{1-s}}\, 
{\frac{\partial \phi(s,\rho)}{\partial \rho}}\big|_{\rho=0}.
\label{spit3}
\end{equation}
To determine the asymptotic behavior of $E[M_n]$ for large $n$, we need to know the
behavior of $h(s)$ near its principal singularity $s=1$. It then follows from Eq. (\ref{spit3})
that we need to know the precise behavior of the function $\phi(s,\rho)$ near $s=1$ and $\rho=0$.
Below we analyse these asymptotic behaviors separately for two cases : (i) For density functions
with a finite second moment, so that ${\hat f}(k) \to 1- \sigma^2k^2/2 +O(k^4)$ as $k\to 0$ where
$\sigma^2$ is the variance of the jump lengths and (ii) for L\'evy flights where the
jump lengths are power law distributed so that ${\hat f}(k) \to 1- |ak|^\mu +O(k^2)$ as $k\to 0$
where $a$ is a microscopic length and $1< \mu\le 2$.

\subsection{ Jump lengths with a finite variance}

In this case, ${\hat f} (k) = 1- \sigma^2 k^2/2 + O(k^4)$ as $k\to 0$. To analyse $\phi(s,k)$ near
$s=1$ and $\rho=0$, it is first necessary to extract the most singular part of $\phi(s,k)$
near $s=1$ and $\rho=0$. To do this, we first rewrite
\begin{equation}
\ln \left(1- s{\hat f}(k)\right) = \ln\left(1-s\left(1-\frac{1}{2}\sigma^2k^2\right)\right)
+ \ln\left( \frac{1-s {\hat f}(k)}{1- s\left(1-\frac{1}{2}\sigma^2k^2\right)}\right).
\label{ana1}
\end{equation} 
We next substitute Eq. (\ref{ana1}) in the expression for $\phi(s,\rho)$ in Eq. (\ref{spit2}) and
subsequently perform the first integral using the following identity\cite{GR}
\begin{equation}
\int_0^{\infty} \frac{\ln\left(1-s+s\sigma^2k^2/2\right)}{\rho^2 +k^2}\, dk=\frac{\pi}{\rho}\, 
\ln\left(\sqrt{1-s}+ \sigma \rho \sqrt{s/2}\right).
\label{iden1}
\end{equation}
This gives
\begin{equation}
\phi(s,\rho)= \frac{1}{\left[\sqrt{1-s}+\sigma\rho \sqrt{s/2}\right]}\, \exp\left[-\frac{\rho}{\pi} 
\int_0^{\infty}
\frac{dk}{\rho^2+k^2}\, \ln\left(\frac{1-s {\hat f}(k)}{1-s + s\sigma^2k^2/2}\right)\right].
\label{ana2}
\end{equation}
The extraction of the most singular part near $s=1$ and $\rho=0$ gives
\begin{equation}
\phi(s,\rho)\approx \frac{1}{\left[\sqrt{1-s}+\sigma\rho/\sqrt{2}\right]}\,
\exp\left[-\frac{\rho}{\pi}
\int_0^{\infty}
\frac{dk}{k^2}\, \ln\left(\frac{1- {\hat f}(k)}{\sigma^2k^2/2}\right)\right],
\label{ana3}
\end{equation}
where $\approx$ in Eq. (\ref{ana3}) means, in the strict mathematical sense, the
following identity,
\begin{equation}
\lim_{s\to 1, \rho\to 0}\frac{1}{\rho}\ln\left[\left(\sqrt{1-s}+\sigma 
\rho\sqrt{s/2}\right)\phi(s,\rho)\right]=-\frac{1}{\pi}
\int_0^{\infty}
\frac{dk}{k^2}\, \ln\left(\frac{1- {\hat f}(k)}{\sigma^2k^2/2}\right).
\label{ana3a}
\end{equation}
Taking the derivative with respect to $\rho$ and putting $\rho=0$ one gets from Eqs. (\ref{ana3})
and (\ref{spit3}), near $s=1$
\begin{equation}
h(s)= \frac{\sigma}{\sqrt{2}} \frac{1}{(1-s)^{3/2}} +\frac{1}{\pi (1-s)}\int_0^{\infty} 
\frac{dk}{k^2} \ln\left[\frac{2}{\sigma^2}\left(\frac{1-{\hat f}(k)}{k^2}\right)\right] 
+O\left(\frac{1}{\sqrt{1-s}}\right).
\label{ana4}
\end{equation}

Noting that the singular behavior $h(s)\sim (1-s)^{-\beta}$ ($\beta>0$) of the generating function
near $s=1$ translates into the estimate $E[M_n]\approx n^{\beta-1}/\Gamma(\beta)$ for large $n$, we get
from Eq. (\ref{ana4}) the following exact asymptotic behaviors of the expected maximum,
\begin{equation}
E[M_n]= \sigma \sqrt{\frac{2n}{\pi}} + \frac{1}{\pi}\int_0^{\infty} \frac{dk}{k^2} 
\ln\left[\frac{2}{\sigma^2}\left(\frac{1-{\hat f}(k)}{k^2}\right)\right] + O(n^{-1/2}).
\label{ana5}
\end{equation}
Dividing by $\sigma$ yields our main result announced in Eq. (\ref{asymp1}) with
an exact expression for $\gamma$ given in Eq. (\ref{c1}). As a check, we find 
that for the exponential density, $f(z)= e^{-|z|}/2$, i.e., with ${\hat f}(k)= 1/(1+k^2)$,
the formula in Eq. (\ref{c1}) yields $\gamma=-1/\sqrt{2}$, in agreement with the
direct solution in Eq. (\ref{sol3}). A few other cases where the integral in Eq. (\ref{c1})
can be performed explicitly are listed in Eqs. (7)-(10). For the case of uniform
density, i.e., $f(z)=1/2$ for $-1\le z\le 1$ (and $0$ otherwise), we thus obtain
an exact closed form expression for $\gamma$ whose numerical value is in agreement
with the result obtained by Coffmann et. al. by a different method\cite{CFFH}.
Our result is evidently more general and holds for arbitrary symmetric jump
density function. In appendix-A, we will prove that $\gamma$ given by Eq. (\ref{c1})
is always negative for arbitrary $f(z)$. 

\subsection{L\'evy distributed jump lengths}

We now consider the case when the second moment of the jump distribution diverges. In particular,
we consider L\'evy jumps such that the Fourier transform of the density function behaves as
${\hat f}(k) = 1- |ak|^{\mu}+O(k^2)$ for small $k$ with $1<\mu \le 2$. This indicates
that in real space the steps lengths have an algebraic tail, $f(x)\sim |x|^{-1-\mu}$ for large $|x|$. 
Note that the Spitzer's formula in Eq. (\ref{spit2}) is still valid for such processes. However,
as we will see, the asymptotic behavior of $E[M_n]$ is quite different from the case where
$\sigma^2$ is finite.

We proceed as in the previous subsection by extracting the most singular behavior of
$\phi(s,\rho)$ near $s=1$ and $\rho=0$. We first rewrite, for $k\ge 0$,
\begin{equation}
\ln \left(1- s{\hat f}(k)\right) = \ln\left(1-s\left(1-(ak)^{\mu}\right)\right)
+ \ln\left( \frac{1-s {\hat f}(k)}{1- s\left(1-(ak)^{\mu}\right)}\right).
\label{lan1}
\end{equation}
We next substitute Eq. (\ref{lan1}) in the expression for $\phi(s,\rho)$ in Eq. (\ref{spit2}).
This gives
\begin{equation}
\phi(s,\rho)= \exp\left[-I_1(s,\rho)-I_2(s,\rho)\right],
\label{lan2}
\end{equation}
where 
\begin{eqnarray}
I_1(s,\rho)&=& \frac{\rho}{\pi}\int_0^{\infty} \frac{dk}{\rho^2+k^2} \ln\left(1-s+s (ak)^{\mu}\right) 
\nonumber \\
&=& \frac{1}{2}\ln(1-s) +\frac{1}{\pi}\int_0^{\infty} \frac{du}{1+u^2} \ln\left[1+ \frac{s}{1-s}
(a\rho u)^{\mu}\right],
\label{lan3}
\end{eqnarray}
and
\begin{equation}
I_2(s,\rho)= \frac{\rho}{\pi}\int_0^{\infty} \frac{dk}{\rho^2+k^2} \ln\left[\frac{1-s {\hat f}(k)}{1-s 
+ s (ak)^{\mu}}\right].
\label{lan4}
\end{equation} 

Taking the derivative with respect to $\rho$ and keeping only the leading singular terms near $s=1$
and $\rho=0$, we get
\begin{equation}
\frac{\partial \phi(s,\rho)}{\partial \rho}\big|_{\rho\to 0}\approx -\frac{1}{\pi \sqrt{1-s}}
\int_0^{\infty} 
\frac{dk}{k^2} \ln\left( \frac{1-s {\hat f}(k)}{(ak)^{\mu}} \right) -
\frac{\mu a^{\mu} \rho^{\mu-1}}{\pi \sqrt{1-s}}
\int_0^{\infty} \frac{u^{\mu}\,du}{(1+u^2)(1-s + (a\rho u)^{\mu})}.
\label{lan5}
\end{equation}
The second integral on the right hand side can further be simplified by first making a change of 
variable
$\rho u= y$ and subsequently taking the limit $\rho\to 0$. The resulting integral can be performed 
in closed form. Putting everything together, we find the leading singular behavior of 
$h(s)$ near $s=1$ and $\rho=0$,
\begin{equation}
h(s) =\sum_{n=0}^{\infty} s^n E[M_n]= \frac{a B(1/\mu, 1-1/\mu)}{\pi (1-s)^{1+1/\mu}} + 
\frac{1}{\pi(1-s)}\int_0^{\infty} \frac{dk}{k^2} \ln\left(\frac{1-s {\hat f}(k)}{(ak)^{\mu}}\right) 
+ O\left(\frac{1}{(1-s)^{1/\mu}}\right). 
\label{lan6}
\end{equation}
where $B(x,y)$ is the standard Beta function.
Subsequently, one obtains the following large $n$ behavior of $E[M_n]$,
\begin{equation}
E[M_n] = \frac{a\mu \Gamma\left(1-\frac{1}{\mu}\right)}{\pi} n^{1/\mu} + \frac{1}{\pi}\int_0^{\infty} 
\frac{dk}{k^2} 
\ln\left(\frac{1- {\hat f}(k)}{(ak)^{\mu}}\right) + O(n^{1/\mu-1}).
\label{lan7}
\end{equation}
Note that for $\mu=2$ and $a= \sigma/\sqrt{2}$, Eq. (\ref{lan7}) reduces to Eq. (\ref{ana5}), as it 
should. The dimensionless expected maximum is obtained by dividing Eq. (\ref{lan7}) by the microscopic 
length $a$ and one arrives at the result in Eq. (\ref{asymp2}) with the constant $\gamma$ given by
the exact formula in Eq. (\ref{c2}). 

\section{Appearance of the constant $\gamma$ in the thermodynamic limit}

In the previous section, we have demonstrated how the constant $\gamma$ appears as the
leading correction term to the asymptotic $\sqrt{n}$ behavior of the expected maximum
$E[M_n]$ for large $n$. In this section we show how $\gamma$ appears as the leading term
in the large $n$ limit if one considers the difference $\left(E[M_n]-E[|x_n|]\right)/\sigma$,
where $E[|x_n|]$ is the expected absolute end to end distance of the walker after $n$ steps.

The calculation of $E[|x_n|]$ is relatively straightforward compared to that of $E[M_n]$.
We start with the random walk in Eq. (\ref{evol1}) where the jump density $f(\xi)$ is
symmetric and has a finite second and fourth moment, $\sigma^2=\int_{-\infty}^{\infty} x^2\,f(x)dx$
and $\mu_4= \int_{-\infty}^{\infty} x^4 \, f(x) dx$. Let $P_n(x)$ be the probability
density for the particle to be between $x$ and $x+dx$ after $n$ steps, starting from $x=0$
at $n=0$. Using the Markov property of the walk, it is easy to see that $P_n(x)$ satisfies the 
recursion relation
\begin{equation}
P_n(x) = \int_{-\infty}^{\infty} P_n(y) f(x-y) dy,
\label{frecur1}
\end{equation}
starting from the initial condition, $P_0(x)= \delta(x)$. Note that unlike the recursion relation
in Eq. (\ref{recur2}), the limits of the integral in Eq. (\ref{frecur1}) are respectively
$-\infty$ and $+\infty$ and hence Eq. (\ref{frecur1}) can be easily solved by taking
the Fourier transform, ${\tilde P}_n(k)=\int_{-\infty}^{\infty} P_n(x) e^{i\, k\, x}$.
One gets ${\tilde P}_n(k) = {\left[{\hat f}(k)\right]}^n$. Hence the solution, valid for all
$n\ge 0$, is obtained from the inverse Fourier transform,
\begin{equation}
P_n(x) = \frac{1}{2\pi} \int_{-\infty}^{\infty} {\left[{\hat f}(k)\right]}^n\, e^{-i\, k\, x}\, dk .
\label{fsol1}
\end{equation}

The large $n$ behavior of $P_n(x)$ can then be obtained by a straightforward
scaling analysis of
the integral in Eq. (\ref{fsol1}). In the large $n$ limit, the most important contributions
to the integral will come from the neighbourhood of $k=0$ where one can expand
${\hat f}(k)= 1- \sigma^2 k^2/2 + \mu_4 k^4/24 + O(k^6)$. This gives, near $k=0$,
\begin{equation}
\ln \left[{\hat f}(k)\right]= -\frac{\sigma^2}{2} k^2 + \frac{\mu_4-3\sigma^4}{24} k^4 + O(k^6)
\label{fsol2}
\end{equation}
Substituting in Eq. (\ref{fsol1}) and rescaling $k\sqrt{n}=q$ and $y=x/\sqrt{n}$ one gets
\begin{eqnarray}
P_n(x)&=& \frac{1}{\sqrt{n}}\int_{-\infty}^{\infty} \frac{dq}{2\pi} \exp\left[-\frac{\sigma^2q^2}{2} +
\frac{(\mu_4-3\sigma^4) q^4}{24 n} +O(q^6/n^2)\right] e^{-i\, q\, y} \nonumber \\
&=& \frac{1}{\sqrt{n}}\int_{-\infty}^{\infty} \frac{dq}{2\pi}e^{-\sigma^2 q^2/2 -i\,q\,y}\left[1+
\frac{(\mu_4-3\sigma^4) q^4}{24 n}+ O(q^6/n^2)\right],
\label{fsol3}
\end{eqnarray}
where we have expanded the exponential for large $n$ holding $y=x/\sqrt{n}$ fixed. 
Using this probability density in Eq. (\ref{fsol3}), it is then easy to calculate
the expectation $E[|x_n|]$ and one gets as $n\to \infty$,
\begin{equation}
E[|x_n|] = \sigma \sqrt{\frac{2n}{\pi}} - \frac{1}{12 
\sqrt{2\pi}}\left[\frac{\mu_4-3\sigma^4}{\sigma^3}\right] \frac{1}{\sqrt{n}} + O(n^{-3/2}),
\label{fsol6}
\end{equation}
in agreement with the result obtained by
Petrov\cite{Petrov} using a somewhat different method. 

Comparing Eq. (\ref{fsol6}) with Eq. (\ref{ana5}), one sees that the leading 
$\sqrt{n}$ term in both $E[|x_n|]$ and $E[M_n]$ have the same coefficient. However,
while the next subleading term in $E[M_n]$ is a constant, the one in $E[|x_n|]$
decays as $n^{-1/2}$ for large $n$. Taking the difference of Eq. (\ref{ana5}) and
Eq. (\ref{fsol6}) and dividing by $\sigma$, we obtain Eq. (\ref{xn2}). Thus, as
$n\to \infty$,
the difference between the two quantities (scaled by $\sigma$) approaches a dimensionless 
constant $\gamma$ given by Eq. (\ref{c1}). Hence it is possible to observe the constant
$\gamma$ even in the thermodynamic limit provided one looks at the difference between
two observables.

\section{Conclusion}

In this paper, we have studied analytically the finite size corrections to the asymptotic
large $n$ behavior of the expected maximum $E[M_n]$ of a one dimensional random walk of $n$ steps with 
arbitrary,
symmetric jump distribution. While the leading $\sqrt{n}$ behavior is universal and easy
to understand by extrapolating the result of continuous time Brownian motion, the 
leading finite size correction term turns out to be a nonuniversal constant $\gamma$ which
is nontrivial. In this paper, we have presented an exact formula for this constant, valid
for arbitrary symmetric jump distribution.
We have also generalized our results to the case of L\'evy processes.

We have also demonstrated how this constant appears even in the $n\to \infty$ limit
as the leading behavior of the difference $E[M_n]-E[|x_n|]$, where $x_n$ is the
position of the walker after $n$ steps. As a nice application, we considered 
a Rouse polymer chain consisting of $n$ monomers at thermal equilibrium
and showed that the difference between the expected positive excursion
along any direction (say the $x$ direction) and the expected absolute value of the
end to end distance of the chain approaches a negative constant in the thermodynamic limit
\begin{equation}
\frac{E[M_n]-E|x_n|}{\sigma}\to \frac{\zeta(1/2)}{\sqrt{2\pi}}= -0.58259\dots.
\label{rouse2}
\end{equation}
where $E$ denotes the thermal average and $\sigma=1/\sqrt{\beta\kappa}$ ($\beta$ being the inverse 
temperature and $\kappa$ being the spring constant of the chain). The result in Eq. (\ref{rouse2})
is nontrivial and somewhat counterintuitive. 

There are several possible extensions of this work. In this paper, we have only considered
the finite size behavior of the expected maximum. It would be nice to extend this finite size
study to higher moments
of the maximum, or even to the full distribution of the maximum. The distribution 
of the maximum of a set of correlated random variables is a subject of current interest
and several papers have recently studied the distribution of the maximum or related objects for 
the continuous time Brownian motion, in the context of fluctuating interfaces\cite{RCPS,MC1,GHPR}
and also in the context of convex polygons and queueing theory\cite{KM}. For discrete
step random walks, the continuous time results (suitably rescaled) will provide
the leading asymptotic behavior of the distribution for large $n$. 
It would be interesting to see if the method presented in this paper can
be used to investigate the finite size effect in the maximum distribution in these systems.

It would also be interesting to extend our results to higher dimensions. For example, is there
a leading constant correction term to $E[M_n]$ for large $n$, where $M_n$ is the 
maximal radial distance from the origin of a random walker of $n$ steps in $d$-dimensions?
If so, it would be interesting to calculate this constant. For this, one needs to develop
a higher dimensional analogue of the Pollaczek-Spitzer formula which would be interesting in its own 
right.

\appendix

\section{Proof that $\gamma< 0$}

In this appendix we show that the constant $\gamma$ that appears in Eq. (\ref{c1}) or Eq. (\ref{c2})
is negative for arbitrary symmetric jump density function $f(x)$. For convenience, we
first rewrite $\gamma= C/\sigma$ where 
\begin{equation}
C = \frac{1}{\pi} \int_0^{\infty} \frac{dk}{k^2} \ln\left[\frac{1-{\hat f}(k)}{\sigma^2k^2/2}\right],
\label{a1.1}
\end{equation}
where ${\hat f}(k)=\int_{-\infty}^{\infty} f(x) e^{ikx} dx$. We assume that $f(x)$ is
symmetric with a finite second moment $\sigma^2$. Clearly, the argument inside the
logarithm in Eq. (\ref{a1.1}) is positive. To prove that $C<0$, it is sufficient
to prove that the argument $[1-{\hat f}(k)]/{\sigma^2k^2/2}$ inside the logarithm 
is less than $1$. For symmetric $f(x)$ one can write ${\hat f}(k)= 2\int_0^{\infty} f(x) \cos (kx) dx$.
The next step is to write the identity
\begin{equation}
\frac{\left(1-{\hat f}(k)\right)}{{\sigma^2 k^2}/2}= 1- \frac{4}{k^2\sigma^2}\int_0^{\infty} f(x) 
\left[\cos (kx)-1 + \frac{k^2x^2}{2}\right] dx,
\label{a1.2}
\end{equation}
which can be proved by carrying out the integration on the right hand side explicitly and
using the definition $\sigma^2 = 2\int_0^{\infty} x^2 f(x) dx$. To prove that the left hand side
of Eq. (\ref{a1.2}) is less than $1$ we need just to prove that the second term on the right hand side 
is positive. Using the elementary inequality, $\cos(kx)-1+ k^2x^2/2 \ge 0$ for all $x$ and
the fact that $f(x)\ge 0$, it follows immediately that indeed the second term is positive.
Thus one has
\begin{equation}
\frac{\left(1-{\hat f}(k)\right)}{{\sigma^2 k^2}/2} < 1 ,
\label{a1.3}
\end{equation}
which then proves that $C< 0$ and hence $\gamma < 0$. A similar proof can be easily
constructed to show that for L\'evy processes as well $\gamma$ as given in Eq. (\ref{c2})
satisfies the inequality $\gamma < 0$.   

\section{Proof of an identity}

In this appendix we prove the following identity valid for all $1< \mu\le 2$,
\begin{equation}
I= \frac{1}{\pi}\int_0^{\infty} \frac{dk}{k^2} \ln
{\left[\frac{1-e^{-k^{\mu}}}{k^{\mu}}\right]}=\frac{\zeta(1/\mu)}{(2\pi)^{1/\mu} \sin (\pi/{2\mu})},
\label{a2.1}
\end{equation}
where $\zeta(z)=\sum_{m=1}^{\infty} m^{-z} $ is the Riemann zeta function which is usually convergent  
for $z>1$. However, it is possible to analytically continue $\zeta(z)$ for $z<1$\cite{AS} and one 
defines
\begin{equation}
\zeta(z) = \lim_{n\to \infty} \left[\sum_{m=1}^n m^{-z} - \frac{n^{1-z}}{1-z}\right]. 
\label{a2.2}
\end{equation}

We first make a change of variable $k^{\mu}=x$ in the integral in Eq. (\ref{a2.1}). This gives 
\begin{equation}
I= \frac{1}{\mu \pi}\int_0^{\infty} dx x^{-1-1/\mu} \ln\left(\frac{1-e^{-x}}{x}\right).
\label{a2.3}
\end{equation}
Next we make one integration by parts and use the fact that $\mu>1$. This yields
\begin{equation}
I= \frac{1}{\pi} \int_0^{\infty} \left[\frac{1}{e^x -1} -\frac{1}{x}\right] x^{-1/\mu} dx.
\label{a2.4}
\end{equation}
Note that each integral on the right hand side of Eq. (\ref{a2.4}) is separately divergent near
$x=0$,
though their difference is convergent. To make progress, we introduce a small cut-off $\epsilon$
and define
\begin{equation}
I(\epsilon)= \frac{1}{\pi} \int_0^{\infty} \left[\frac{1}{e^x -1 +\epsilon} 
-\frac{1}{x+\epsilon}\right]x^{-1/\mu} dx
= I_1(\epsilon) - I_2(\epsilon) . 
\label{a2.5}
\end{equation}
Eventually we are intersted in $I=I(0)$. The reason behind introducing this additional 
cut-off is so that the two integrals will be separately convergent for finite $\epsilon$ and
then after performing the two separate integrals, we will eventually take the $\epsilon\to 0$
limit. 

The first integral can be written as
\begin{equation}
I_1(\epsilon)= \frac{1}{\pi} \int_0^{\infty} \frac{x^{-1/\mu}dx}{e^x-1+\epsilon}= \frac{1}{\pi} 
\Gamma\left(1-\frac{1}{\mu}\right) \Phi\left(1-\epsilon, 1-\frac{1}{\mu}, 1\right),
\label{a2.6}
\end{equation}
where $\Phi(z,s,v)= \sum_{n=0}^{\infty} (v+n)^{-s} z^n $ is the Lerch function\cite{GR}.
The second integral is elementary and can also be performed exactly to give
\begin{equation}
I_2(\epsilon)= \frac{1}{\pi} \int_0^{\infty} \frac{x^{-1/\mu}\,dx}{x+\epsilon}= 
\frac{{\epsilon}^{-1/\mu}}{\pi} B\left(1-\frac{1}{\mu}, \frac{1}{\mu}\right),
\label{a2.7}
\end{equation}
where $B(x,y)=\Gamma(x)\Gamma(y)/\Gamma(x+y)$ is the standard Beta function. Putting these together, we 
then have
\begin{equation}
I(\epsilon)= \frac{1}{\pi}\Gamma\left(1-\frac{1}{\mu}\right)\left[\Phi\left(1-\epsilon, 
1-\frac{1}{\mu}, 1\right)-\Gamma\left(\frac{1}{\mu}\right) {\epsilon}^{-1/\mu}\right].
\label{a2.8}
\end{equation}

Now, the tricky part is to take the $\epsilon\to 0$ limit in Eq. (\ref{a2.8}). To do this, we will
make use of the following asymptotic behavior of $\Phi(z,s,v)$ as $z\to 1$\cite{MOS}    
\begin{equation}
\Phi(z,s,v) = \Gamma(1-s) [-\ln z]^{s-1} z^{-\alpha} + \zeta(s,\alpha)
\label{a2.9}
\end{equation}
where $\zeta(s,\alpha)= \sum_{n=0}^{\infty} (n+\alpha)^{-s}$. Putting $z=1-\epsilon$
in Eq. (\ref{a2.9}) and expanding for $\epsilon$, one then finds that as $\epsilon\to 0$, the
two leading order terms are given by
\begin{equation}
\Phi\left(1-\epsilon, 1-\frac{1}{\mu}, 1\right) \to \Gamma\left(\frac{1}{\mu}\right) 
{\epsilon}^{-1/\mu}
+\zeta\left(1-\frac{1}{\mu},1 \right) +O(\epsilon^{1-1/\mu}).
\label{a2.10}
\end{equation}
Substituting this behavior in Eq. 
(\ref{a2.8}) we get
\begin{equation}
I =I(\epsilon\to 0)= 
\frac{1}{\pi}\Gamma\left(1-\frac{1}{\mu}\right)\zeta\left(1-\frac{1}{\mu},1\right)=
\frac{1}{\pi}\Gamma\left(1-\frac{1}{\mu}\right)\zeta\left(1-\frac{1}{\mu}\right),
\label{a2.11}
\end{equation}
where we have used the fact that $\zeta(z,1)= \zeta(z)$. One can further rewrite Eq. (\ref{a2.11})
by using the identity\cite{GR},
\begin{equation}
\zeta(z)= 2 (2\pi)^{z-1} \sin(\pi z/2) \Gamma(1-z) \zeta(1-z).
\label{a2.12}
\end{equation}
Using this identity in Eq. (\ref{a2.11}) one readily arrives at the final result in Eq. (\ref{a2.1}).

In particular, note that the result in Eq. (\ref{a2.1}) is valid even for $\mu=2$. In that case
one gets, $I (\mu=2)= \zeta(1/2)/\sqrt{\pi}$. Note that when the jump lengths are Gaussian distributed
as in Eq. (\ref{special1}), one has ${\hat f}(k)= \exp[-k^2\sigma^2/2]$. Hence, from Eq. (\ref{c1})
\begin{equation} 
\gamma= \frac{1}{\pi \sqrt{2}}\int_0^{\infty} \frac{dk}{k^2} \ln 
\left[\frac{1-e^{-k^2}}{k^2}\right]=\frac{I(\mu=2)}{\sqrt{2}}= 
\frac{\zeta(1/2)}{\sqrt{2\pi}}=-0.58259\dots,
\label{a2.13}
\end{equation} 
thus proving the result in Eq. (\ref{special1}). The numerical value of $\zeta(1/2)=-1.46035$ can be 
obtained
to arbitrary precision using Mathematica and was used in Eq. (\ref{a2.13}).

\vspace{0.5cm}

{\em Note added in proof:} We thank R.M. Ziff for pointing out that the constant $c=\gamma\sigma=-0.297952..$
for the uniform case also appeared in an apparently unrelated
three dimensional trapping problem first
studied in [Ziff R. M., {\em  flux to a trap}, 1991 J. Stat. Phys. {\bf 65} 1217]
where it was evaluated by numerically iterating a set of recurrence relations.
Our Eq. (11) provides an exact expression of this constant.

\end{document}